\RequirePackage{lineno}
\documentclass[5p]{elsarticle}

\usepackage{epsfig}
\usepackage{amssymb}
\usepackage{bm}
\usepackage{graphicx,color}
\usepackage{dcolumn}
\usepackage{longtable}
\usepackage{epstopdf}
\usepackage{csquotes}
\usepackage{amsmath}
\usepackage{physics}
\usepackage[colorlinks=true,backref=page]{hyperref}
\usepackage{caption}
\usepackage{subcaption}
\usepackage{placeins} 

\begin{document}

\title{Polarized Neutron Measurements of the Internal Magnetization of a Ferrimagnet Across its Compensation Temperature}

\author[1,2,3]{C. D. Hughes}

\author[1,2,3]{K. N. Lopez\corref{cor1}}
\ead{knlopez@iu.edu}

\author[4]{T. Mulkey}

\author[5]{B. Hill}

\author[5]{J. C. Long}

\author[4]{M. Sarsour}

\author[1,2,3]{M. Van Meter}

\author[1,2,3]{S. Samiei}

\author[1,2]{D. V. Baxter}

\author[1,2,3]{W. M. Snow}

\author[6]{L. M. Lommel}

\author[1,2,3]{M. Luxnat}

\author[7]{Y. Zhang}

\author[7]{C. Jiang}

\author[7]{E. Stringfellow}

\author[7]{P. Zolnierczuk}

\author[7]{M. Frost}

\author[7]{M. Odom}

\cortext[cor1]{Corresponding author.}

\affiliation[1]{organization={Department of Physics, Indiana University},
                addressline={727 E. 3rd St.}, 
                city={Bloomington},
                state={IN},
                statesep={}, 
                postcode={47405}, 
                country={USA}}

\affiliation[2]{organization={Center for Exploration of Energy and Matter},
                addressline={2401 N. Milo B. Sampson Ln.}, 
                city={Bloomington},
                state={IN},
                statesep={}, 
                postcode={47408}, 
                country={USA}}

\affiliation[3]{organization={Indiana University Center for Spacetime Symmetries},
                addressline={727 E. 3rd St.}, 
                city={Bloomington},
                state={IN},
                statesep={}, 
                postcode={47405}, 
                country={USA}}

\affiliation[4]{organization={Department of Physics and Astronomy, 
                            Georgia State University},
                city={Atlanta},
                state={GA},
                statesep={},
                postcode={30303},
                country={USA}}

\affiliation[5]{organization={Department of Physics, 
                            University of Illinois at Urbana-Champaign},
                city={Urbana},
                state={IL},
                statesep={},
                postcode={61801},
                country={USA}}

\affiliation[6]{organization={University of Notre Dame},
                addressline={Holy Cross Dr.},
                city={Notre Dame},
                state={IN},
                statesep={},
                postcode={46556},
                country={USA}}

\affiliation[7]{organization={Oak Ridge National Laboratory},
                city={Oak Ridge},
                state={TN},
                statesep={},
                postcode={37830},
                country={USA}}

\date{\today}

\begin{abstract}

We present the first polarized neutron transmission image of a model Ne\'el ferrimagnetic material, polycrystalline terbium iron garnet (Tb$_{3}$Fe$_{5}$O$_{12}$, TbIG for short), as it is taken through its compensation temperature $T_{comp}$ where the macroscopic magnetization vanishes. Our polarized neutron imaging data and the additional supporting measurements using neutron spin echo spectroscopy and SQUID magnetometry are all consistent with a vanishing internal magnetization at $T_{comp}$.

\end{abstract}

\maketitle

\section{Introduction}

Ferrimagnetic materials contain distinct sublattices of atoms with magnetic moments of different magnitudes and antiferromagnetic coupling among those atoms. The different moment size leads to a net magnetic moment at the unit cell level despite this antiferromagnetic coupling. In general, the temperature dependence of the sublattice magnetizations can be different, so it is possible for the magnetic moment per unit cell of the ferrimagnet to go to zero at a particular temperature that is known as the magnetic compensation temperature ($T_{comp}$). If the gyromagnetic ratios of the two atoms differ, the unit cell will also exhibit a non-zero total angular momentum $\vec{J}$ at $T_{comp}$. Moreover, since the ratio of the orbital and spin components of the total angular momentum may also vary for different atoms, there can remain some nonzero total orbital angular momentum $\vec{L}$ and a nonzero total spin angular momentum $\vec{S}$ at $T_{comp}$ as well. The total angular momentum ($\vec{J}=\vec{L}+\vec{S}$) may vanish at a different temperature $T_{A} \ne T_{comp}$. 

The existence of ferrimagnetism has been known for several decades \cite{Neel48,Geller65,Dionne71}. Textbooks and reviews on the phenomenon describe what is known experimentally, which consists mostly of measurements of the external magnetic field produced by the internal magnetization, and the theory of ferrimagnetism is thought to be reasonably well understood \cite{Bara82,Dionne09Book,Dionne76,Herbst82,Herbst84}. Recently, special aspects of this phenomenon have attracted scientific attention from new directions. If a ferrimagnet possesses magnetic domains at $T_{A}$, for example, the fast domain-wall dynamics predicted for $\vec{J}=0$ antiferromagnetic materials might occur instead in a material with a net magnetic moment, which in turn could enable these domains to be  manipulated with external magnetic fields \cite{Kim2017,Hirata19}. In addition, a ferrimagnet at $T_{comp}$ realizes a very interesting new type of polarized target medium: according to the simple model described above, it can be understood as an ensemble of polarized electrons with no net internal or external magnetic field.  

Measurements characterizing ferrimagnetic phenomena have typically sensed the readily-accessible macroscopic external fields, and applications for these materials typically rely on the net moment. There has been relatively little detailed study of the microscopic magnetic structure of these materials at $T_{comp}$. For applications that rely on a net spin polarization at $T_{comp}$, there remains some concern that the simple mean-field description of the local spin structure could be complicated by disorder near domain walls and other deviations from the simplistic model of the macroscopic sample being a homogeneous periodic replication of ideal unit cells. These possibilities motivate a careful study of the internal magnetic structure of a ferrimagnetic material as it is taken through its magnetic compensation point.
 
 In this paper we present to our knowledge the first images of the internal magnetization of a ferrimagnet taken through $T_{comp}$. With the advent of polarized neutron imaging, one can measure the internal magnetic fields of a material using polarized neutron spin rotation in transmission.
 
 The net electron spin polarization with no net magnetization of a compensated ferrimagnet at $T_{comp}$ makes such systems an excellent choice to search for possible exotic spin-dependent interactions of electrons with matter without an overwhelming spin-dependent background signal from magnetic interactions. Such exotic spin-dependent interactions are predicted in a very wide variety of theories beyond the Standard Model of particles and interactions, including string theory \cite{Leitner64, Weinberg72, Moody84, Dob06, Ade09, Jae10, Leslie2014, Murata15}. Experiments must be optimized for different regimes within the wide spectrum of possible interaction ranges. The range of these interactions is inversely proportional to the mass of the associated exchange boson. One spin-dependent exotic interaction search in preparation \cite{Leslie2014} uses a ferrimagnet attached to a mechanical oscillator with sensitivity at the thermal noise limit to investigate a wide variety of possible exotic spin-dependent interactions with macroscopic range. A recent review \cite{Safronova2018} and a book chapter \cite{Flambaum23} have placed this work in the context of analogous investigations using atomic measurements. 

Slow neutrons are well-suited for such searches \cite{Nico05b, Dubbers11, Pignol2015, Sponar2021, Snow2021}, and several sensitive neutron experiments have been conducted to constrain such interactions \cite{Leeb92, Frank2004, Bae07, Nez08, Ser09, Voronin2009, Pok10, Pie12, Jen14, Afach2015, Lemmel2015, Kamiya2015, Cronenberg2018, Heacock21, Yan13, Lehnert14, Li2016, Lehnert17, Haddock2018a, Haddock2018b, Parnell2020}. The ability of slow neutrons to move through matter with minimal quantum decoherence enables phase-sensitive interferometric measurement methods of various types. Efficient techniques to polarize and analyze neutron beams and manipulate the neutron spin are well-developed.     

A sensitive experimental search for possible exotic spin-dependent interactions of polarized neutrons with polarized electrons can uncover new interactions of Nature inaccessible by other methods. For shorter range interactions, the number of spin-polarized particles that can be brought within the relevant interaction range simply by bringing two macroscopic objects close to each other becomes smaller and smaller. It also becomes more and more difficult to maintain the required control over systematic background effects from magnetic field distortions generated by test mass susceptibility and magnetic impurities. Spin-dependent interactions with short ranges inaccessible to measurements with macroscopic bodies can be searched for using coherent neutron transmission measurements. The neutron optical potential of the matter that the neutron passes through is sensitive to any contribution to the real part of the forward scattering amplitude. These contributions include any spin-dependent interactions and those from the exotic interaction of interest.

The number of possibilities for the form of such exotic interactions beyond the Standard Model of particles and interactions is not as arbitrary as one might suppose. Theoretical work on the possible interactions among point particles \cite{Dob06,Fadeev2019} reveals 16 linearly-independent nonrelativistic potentials from the single exchange of a spin-$0$ or spin-$1$ boson composed from scalar invariants involving the spins, momenta, interaction range, and possible particle couplings between nonrelativistic spin-$1/2$ fermions.  Experimental laboratory constraints on the subset of these possible interactions involving two polarized species are scarce, especially for short-range interactions.  ``In-matter” gravitational torsional fields can be sourced by the polarized electron ensemble. Just as mass curves spacetime in general relativity, the other property that all matter particles possess in addition to mass, namely spin, has long been thought to produce an additional ``twist” of spacetime, geometrically independent of curvature, called torsion \cite{Hehl76,Shapiro02,Hammond02}. To experimentally probe torsion effects at short distances \cite{Carroll94} requires one to probe inside a dense system of polarized particles. Polarized neutron transmission through a magnetically-compensated ferrimagnet is one of the very few known systems in which one can perform such an experimental search.

To use a magnetically-compensated ferrimagnet to search for new interactions, it is essential to extract direct information of the actual magnetic fields inside the sample at $T_{comp}$. Previous work using polarized neutron scattering as an internal probe of ferrimagnetic materials \cite{Louca2009,Bonnet79,Nambu20} did not investigate the magnetic structure at $T_{comp}$.  Polarized neutron imaging can directly measure the line integrals of the internal magnetic fields along the neutron trajectories using polarized neutron spin rotation. Neutron spin echo spectroscopy in transmission can achieve even higher sensitivity to internal fields, albeit with lower spatial resolution. 

The remainder of this paper is organized as follows. Section \ref{sec:SampleFabricationAndMagnetization} describes our sample fabrication and magnetization procedure. Section \ref{sec:SampleCharacterization} summarizes our local characterization studies of TbIG ferrimagnetic samples using SQUID magnetometry and x-ray diffraction. Section \ref{sec:PreviousNeutronWork} provides detail on our custom sample environment and temperature control system. Section \ref{sec:NSEMeasurements} describes our results from the Spallation Neutron Source Neutron Spin Echo spectrometer at Oak Ridge National Laboratory (ORNL). Section \ref{sec:PolNeutronImagingSetup} describes the neutron polarimeter setup on the High Flux Isotope Reactor (HFIR) neutron imaging beamline at ORNL, and Section \ref{sec:PolNeutronSpinRotImages} presents 2D images of the net polarized neutron spin rotation angle through the sample as a function of the sample temperature as it is taken through $T_{comp}$. These images clearly show the magnetization vanishing at $T_{comp}$ and reversing sign above and below $T_{comp}$. We compare the magnetization inferred from this data with a mean field model of ferrimagnetism as applied to TbIG, and we also use this model to infer the polarized electron spin density at $T_{comp}$. We conclude with a description of ongoing data analysis of the polarized neutron images to learn about or constrain the possible presence of ferrimagnetic domains and to search for possible exotic spin-dependent interactions of polarized neutrons and polarized electrons. 

\subsection{Review of ferrimagnetism in TbIG}

TbIG is an intermetallic compound with the magnetic properties of a ferrimagnet. Ferrimagnets exhibit spontaneous alignment of magnetic moments within domains like ferromagnets, but the resulting magnetization is low, since the magnetic moments of the sublattice species oppose each other. Here, the Tb lattice aligns with one Fe lattice and anti-aligns with another (see Sec. \ref{sec:PolNeutronSpinRotImages}). The magnitudes $\mu_{Fe}(T)$ and $\mu_{Tb}(T)$ of the respective moments vary with temperature with a dependence that can be calculated in a mean-field model given sufficient knowledge of the various moment-moment couplings. The magnetic moment of the iron is mostly due to the spin of electrons, but that of the terbium has a significant contribution from the orbital motion of the inner shell 4f electrons.

The moment directions can possess a small canting angle $\theta_{c}$ determined by interatomic forces within the crystal. In this paper we will assume, consistent with our observations, that the canting angle $\theta_{c}$, if present, is a fixed property of the crystal lattice, independent of temperature and applied magnetic field. As the data will show, we see no evidence for a net nonzero canting angle upon averaging over the sample volume.

\section{Sample Fabrication and Magnetization}
\label{sec:SampleFabricationAndMagnetization}

\subsection{Fabrication}

Our polycrystalline rare-earth iron garnet samples are produced using a coprecipitation method \cite{Weisman17,Geselbracht94,Azis18,Khattak76}. Stoichiometric amounts of Re(NO$_3$)$_3$ and FeCl$_3$ are added to deionized water to create 1 molar (M) stock solutions, where Re is the rare earth element of choice. A 6 M solution of NaOH acts as the precipitating agent in the reaction, and the resulting solution containing the precipitant  is brought to a neutral pH with a deionized water wash before being dried at 120$^{\circ}$ C for at least 12 hours. The dried material is subsequently calcined for at least 12 hours at 1200$^{\circ}$C. The material is then ground, pressed, and fired again to increase the purity. 

Powder samples of TbIG were fabricated from freshly synthesized material, pressed into cylindrical samples, and fired in a furnace up to temperatures of 1200$^{\circ}$C. The resulting samples were 19 mm in diameter and 7.4 mm thick, with density 3.8 g/cm$^3$.

\subsection{Magnetization}

Like a ferromagnet, a ferrimagnet is magnetized by exposing it to an external magnetic field intense enough to saturate the magnetization. In our case, this magnetization step was performed at room temperature. The cylindrical TbIG samples were magnetized along the longitudinal axis using a large NMR electromagnet capable of field strengths near 1 Tesla with fractional magnetic field uniformity of better than $10^{-4}$ over the sample volume. The accuracy of alignment of the external magnetic field direction and the axis of the sample was better than 1 mrad. 

\section{Sample Characterization}
\label{sec:SampleCharacterization}

\subsection{X-Ray Diffraction}

\begin{figure}
    \includegraphics[width=\columnwidth]{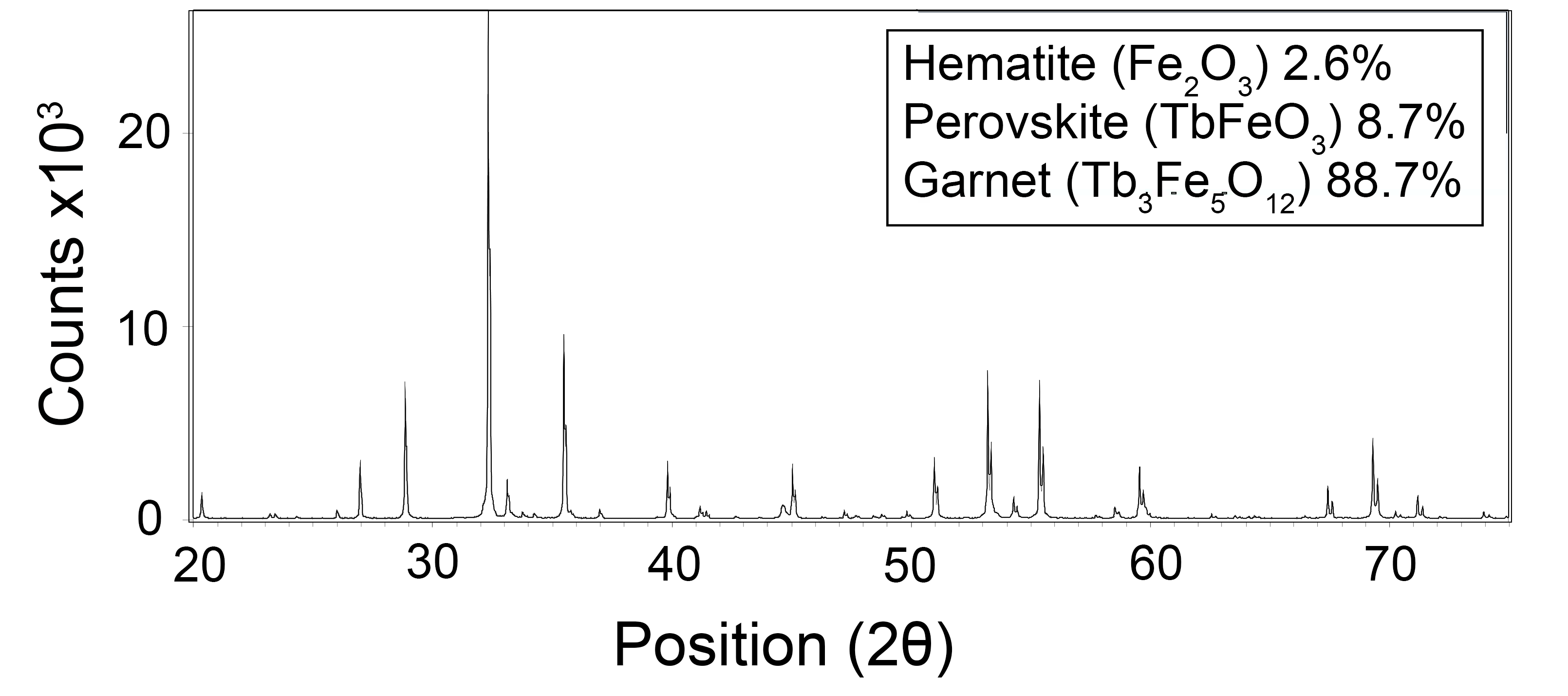}
    \caption{Measured X-ray intensity vs. 2$\theta$ angle. Diffraction pattern is typical of a terbium iron garnet, with additional peaks contributed by contaminants. Results of a Rietveld analysis of this TbIG sample are displayed in the upper right corner.}
   \label{fig:XRD_Rietveld}
\end{figure}

We employed X-ray powder diffraction measurements to help tune the sample preparation process, minimizing the perovskite phase (TbFeO$_{3}$) and hematite (Fe$_{2}$O$_{3}$), the principal observed contaminants. Material was sequestered from the bulk before samples were pressed into pellets. The samples were then analyzed in a Bruker D8 Advance PXRD device using CuK$\alpha$ radiation at the Indiana University Molecular Structure Center. The resulting scans were compared to standards in either the Powder Diffraction File-2 (PDF-2) database or the Inorganic Crystal Structure Database (ICSD). A Rietveld refinement of the data was performed to determine composition fractions. Using this method, we determined the purity of sample Tb082422\_f103 to be 88.7\% Tb$_{3}$Fe$_{5}$5O$_{12}$ (weighted residual profile value of 13.80\% and a $\chi^{2}$ goodness-of-fit of 4.65), with TbFeO$_{3}$ making up the majority of the remainder along with some contribution from Fe$_{2}$O$_{3}$. Figure \ref{fig:XRD_Rietveld} shows the results of a Rietveld analysis of an optimized sample. 

Both of the contaminants can, in principle, contribute to the magnetic signal.  Based on our knowledge of their properties from previous scientific literature, we anticipated that they would not interfere with the interpretation of our experimental data. The orthoferrite TbFeO$_3$ is a weak antiferromagnet with anti-aligned but slightly canted magnetic moments \cite{Treves1965}. The hematite Fe$_{2}$O$_{3}$ is a pure antiferromagnet under 260 K, but undergoes a Morin transition at 260 K into a weak antiferromagnet \cite{Filho2014}. Antiferromagnetic ordering will not contribute to the experimental observables we employ, and we see no evidence for magnetic contributions from these sample contaminants in our data as will be clear below.

\subsection{SQUID Magnetometry}

\begin{figure}
    \centering
    \includegraphics[width=\columnwidth]{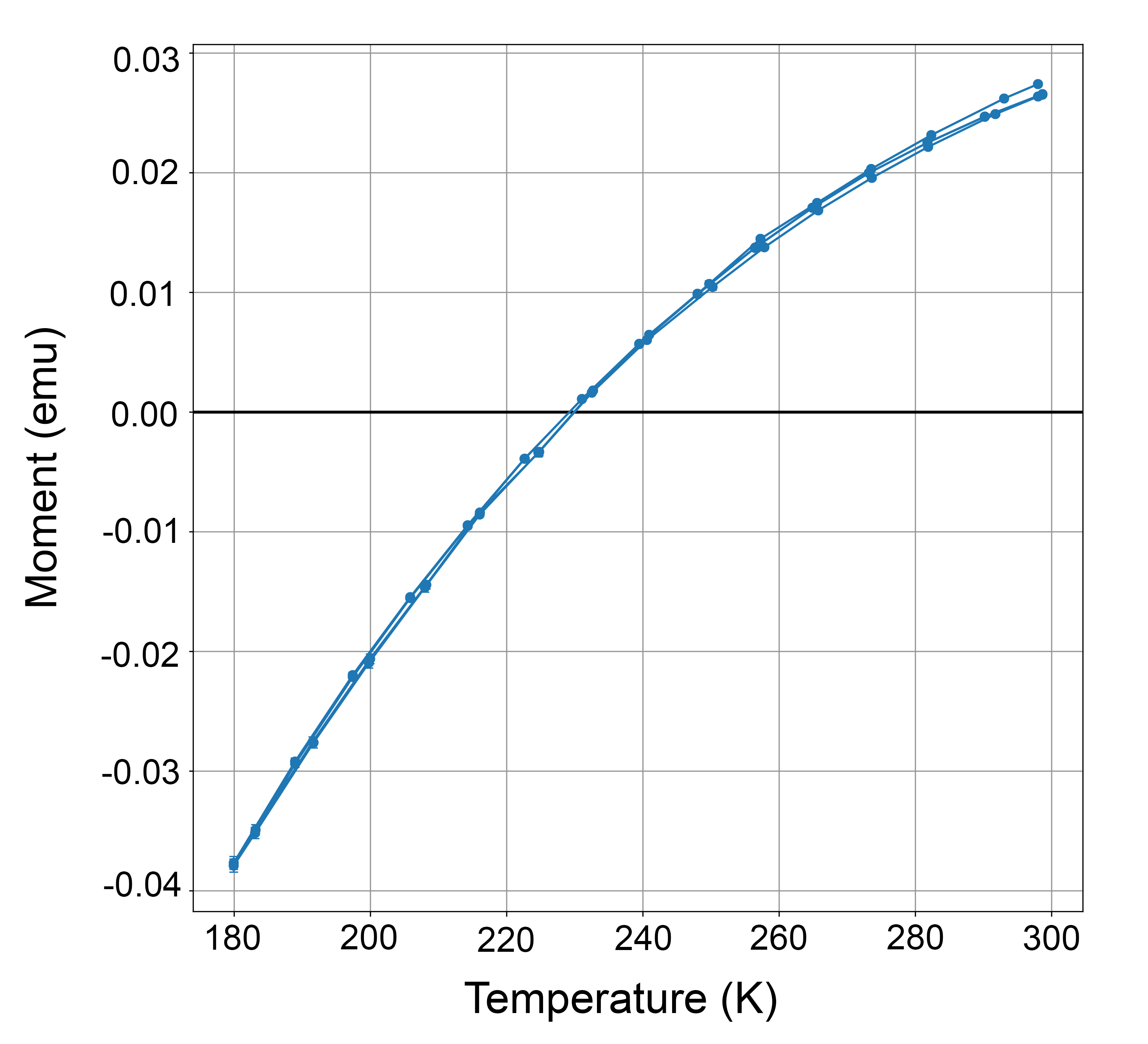}
    \caption{Total remanent magnetic moment (in zero applied field) of a small (0.32 cm diameter) DyIG sample as a function of temperature, as measured in MPMS magnetometer. Data are for several cycles of the temperature between 180 K and 298 K.}
    \label{fig:DyIG_sQUID}
\end{figure}

\begin{figure}
    \centering
    \includegraphics[width=\columnwidth]{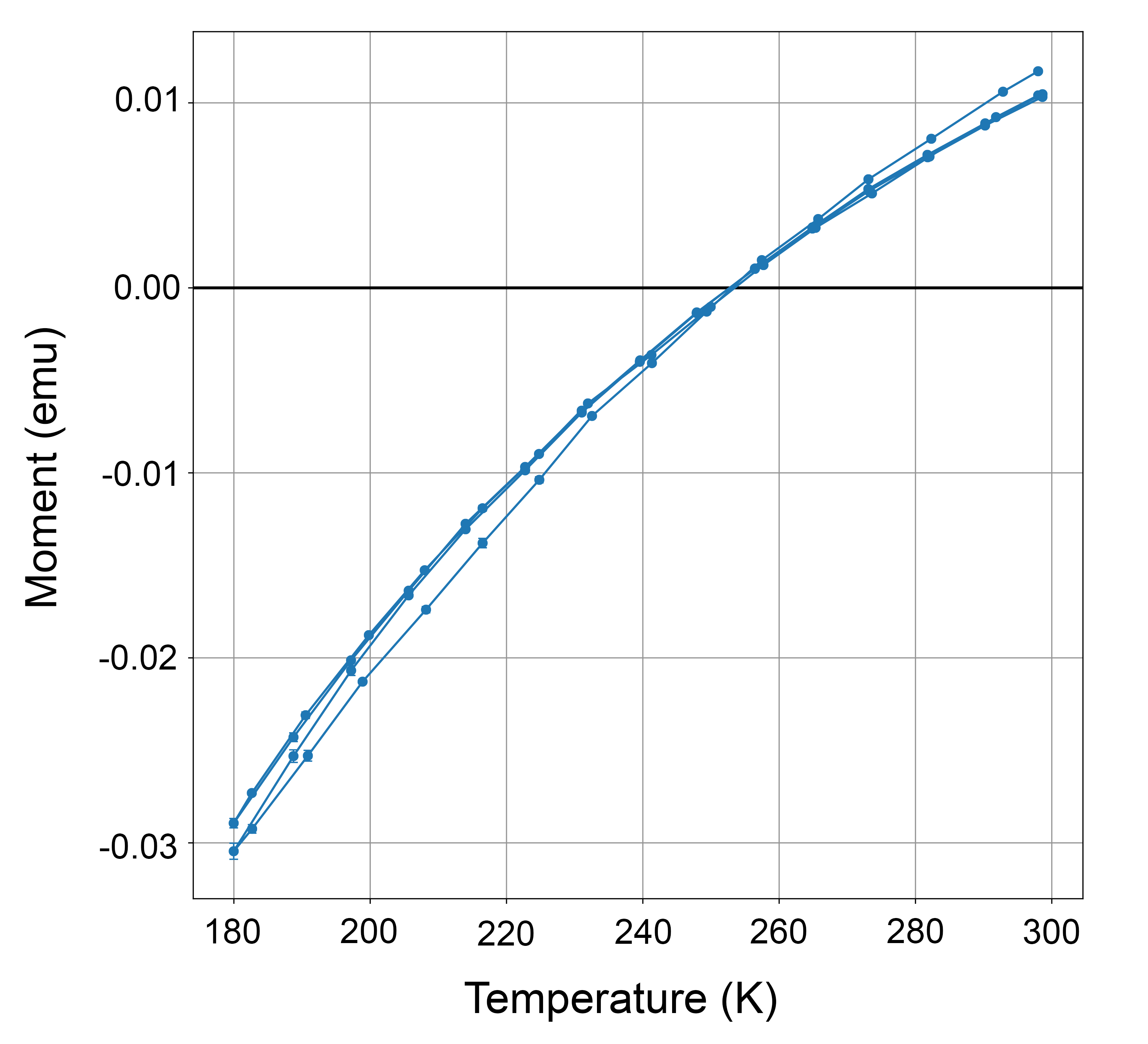}
    \caption{Total remanent magnetic moment (in zero applied field) of a 0.32 cm diameter TbIG sample as a function of temperature, as measured in the MPMS magnetometer. Data are for several cycles of the temperature between 180 K and 298 K. The slight variations between cycles, away from $T_{comp}$, are consistent with a small amount of sample motion from the limited rigidity of the sample mount.}
    \label{fig:TbIG_sQUID}
\end{figure}

We measured the external fields from TbIG and also dysprosium iron garnet (DyIG) samples using a Quantum Design MPMS-XL SQUID system to infer a compensation temperature of  $229.5 \pm  0.5$ K for DyIG (Fig. \ref{fig:DyIG_sQUID}) and $253 \pm 1$ K for TbIG (Fig. \ref{fig:TbIG_sQUID}). These compensation temperatures were reproducible under different sample preparation states and upon thermal cycling, and showed no evidence of hysteresis within measurement errors and instrumentation limitations.
 
\section{Previous Neutron Work}
\label{sec:PreviousNeutronWork}

Few neutron experiments involving TbIG have been performed. Some references to work from more than three decades ago exist \cite{Baazov83,Mand85,Fuess76A,Fuess76B}. More abundant literature is available on yttrium iron garnets (YIG), a popular material used for many years in spintronic and microwave devices. Plant et al. \cite{Plant77} used unpolarized neutrons to explore the dispersion of  spin wave excitations. More recent work using unpolarized \cite{Man17,Shamoto18,Princep17} and polarized \cite{Nambu20,Bonnet79} neutrons has been conducted to further explore spin waves, magnon polarization, and other magnetic properties in YIG. To our knowledge, none of the neutron measurements described below have been previously conducted on TbIG. 

\subsection{Sample Environment Design and Temperature Control for Polarized Neutron Measurements}

\begin{figure}
    \centering
    \includegraphics[width=0.7\columnwidth]{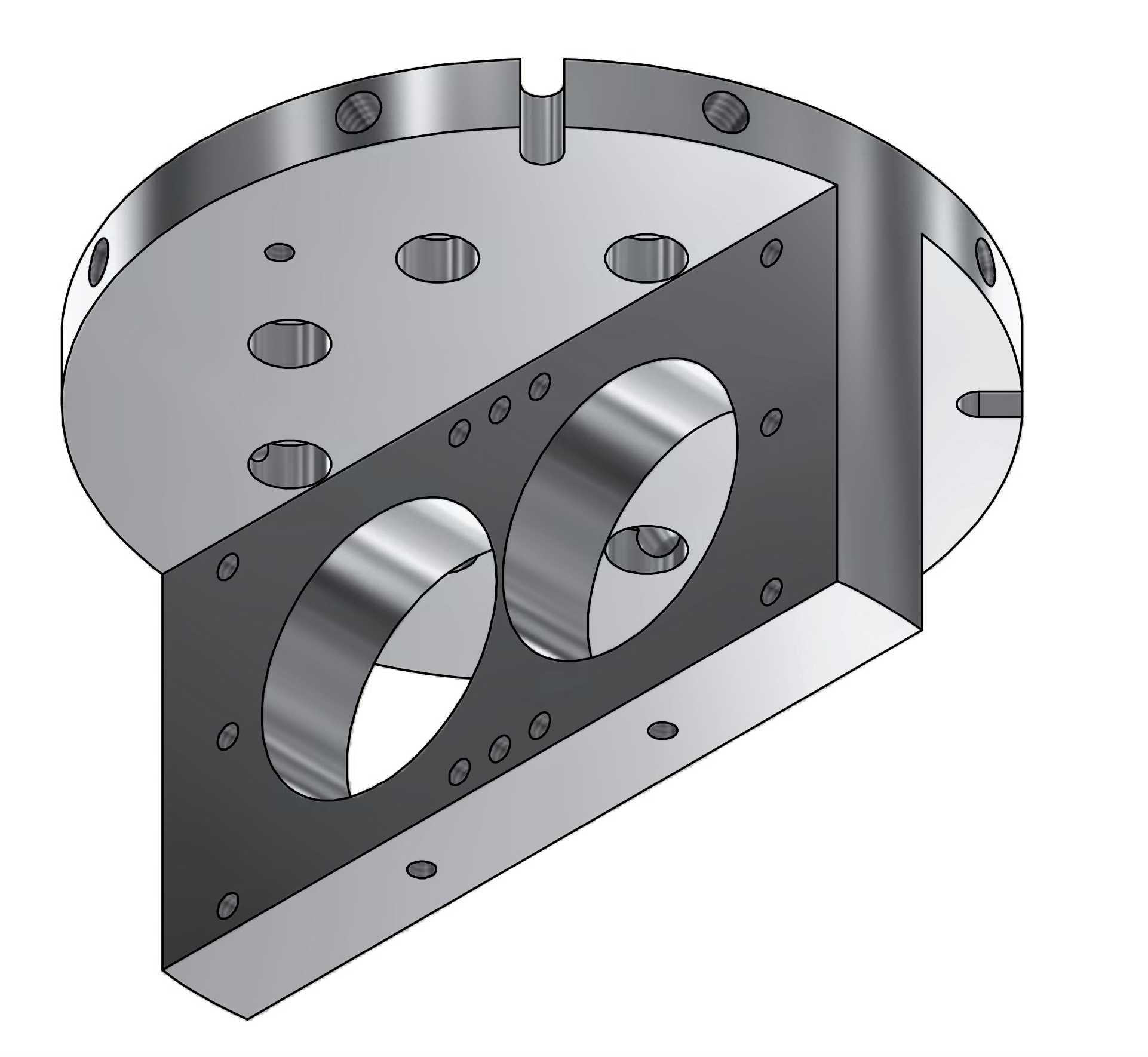}
    \caption{The two-cell sample holder used in the SNS-NSE experiment.}
    \label{fig:T2}
\end{figure}

 The TbIG samples are held in a custom aluminum mount shown in Figure \ref{fig:T2} which allows for two samples to be mounted simultaneously. Aluminum has a high thermal conductivity and also a relatively low neutron scattering and absorption cross section compared to TbIG.  The mount allows for up to four silicon diode sensors to be attached for thermometry. The mount is secured to a copper disk at the end of a cold finger inside of a Cryo Industries Model RC102 sample-in-vacuum continuous flow cryogenic workstation. A thin aluminum radiation shield designed to minimize possible thermal gradients in the sample from nonuniform radiative heating attaches to the outer rim of the mounting disk. Three sapphire plates manufactured by Gavish Sapphire Products hold the samples in place. Sapphire was chosen for this purpose due to its low neutron absorption and scattering cross section and high thermal conductivity. One large sapphire plate remains flush with the sample mount such that the sample side of the plate, the edge of the sample mount, and the sample face are coplanar. The other face of the pellet is secured using a smaller sapphire plate, half the length of the large plate and the same width and thickness. This allows for samples of different thicknesses to be mounted in the two sample cells. Both sapphire plates were held in place with aluminum brackets and brass screws which were scanned for internal magnetic fields from magnetic impurities and found to have no fields larger than 50 nT a few mm away from the sensitive volume of a fluxgate magnetometer. To improve thermal contact area and thereby minimize thermal gradients in the sample, a small amount of Krytox grease was applied to the plate where it makes contact with the pellet. Krytox is a fluorinated grease chosen to avoid the strong neutron scattering that would arise from any grease composed of hydrogenous materials. 

 The body of the cryostat is made of aluminum with four windows on either side of the cubic base. The main vacuum windows of the cryostat were replaced with sapphire windows, also produced by Gavish. The original screws for the retaining ring of the cryostat windows were replaced with brass screws and washers scanned for magnetic impurities using fluxgate magnetometers. A rough vacuum is maintained within the volume of the cryostat using a dry scroll pump.
 
 The cryogenic coolant is liquid nitrogen. The cryogen is allowed to flow into the cryostat at a rate controlled by the dewar pressure and conductance of the transfer line before boiling and venting to atmosphere. The temperature of the system is controlled via PID feedback loop using a Lakeshore Model 331 temperature controller, managed via a slow-control LabVIEW program, the silicon-diode temperature sensors, and a resistive heater embedded in the sample mount at the end of the coldhead. With this scheme, we are able to reliably achieve temperature stability of 100 mK.
 
\section{Neutron Spin Echo Measurements}
\label{sec:NSEMeasurements}

We used the Neutron Spin Echo spectrometer at Oak Ridge National Laboratory's Spallation Neutron Source (SNS-NSE) to investigate the internal magnetization of TbIG as a function of temperature. In transmission mode, the NSE spectrometer can measure the line integral of the internal magnetic field of the sample along the neutron trajectories using the phase shift of the interference fringes of the neutron spin echo signal. NSE can also search for possible components of sample magnetization normal to the neutron momentum transfer $\vec{Q}$ by using the sample itself as the $\pi$ flipper to generate the spin echo pattern (the so-called ``paramagnetic" neutron spin echo mode)~\cite{mezei_fundamentals_2003,keller_neutron_2022,pappas_polarimetric_2008,mezei_principles_1980,chatterji_neutron_2006}. 

\begin{figure}
    \centering
    \includegraphics[clip,trim=0.5cm 1.7cm 0.7cm 2cm,width=\columnwidth]{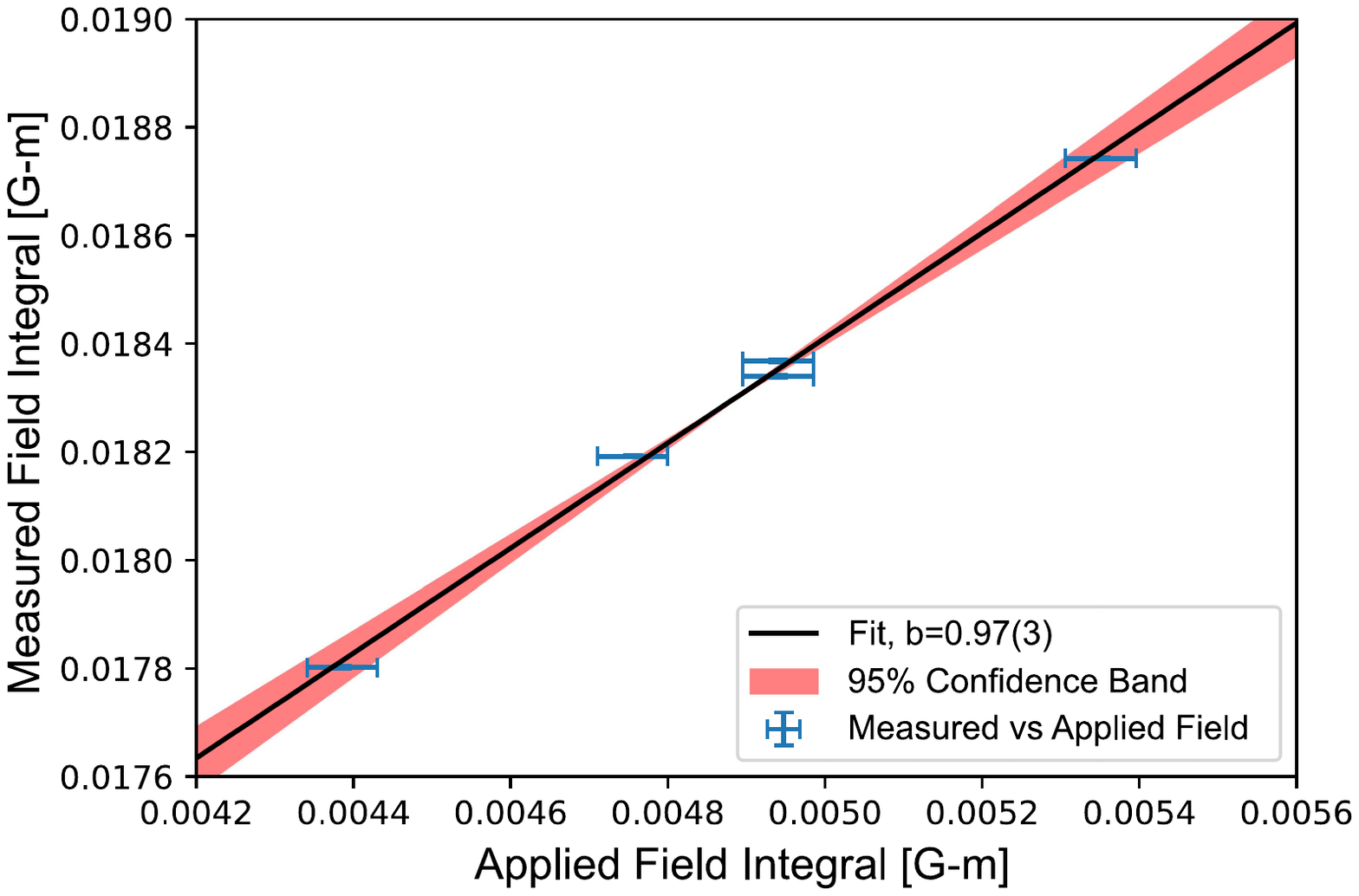}
    \caption{Calibration of the SNS Neutron Spin Echo instrument's phase shift versus a known, applied magnetic field integral in the sample region as measured over a neutron wavelength range between 1.3 meV and 3.2 meV. }
    \label{fig:NSESens}
\end{figure}

Before the measurement, we calibrated the response of the SNS neutron spin echo spectrometer in transmission mode to a known magnetic field integral applied in the sample region (Fig. \ref{fig:NSESens}). This measurement achieved a 100 nT-m field integral sensitivity in a half-hour of beam time with the SNS spallation target power at 1.5 MW.

\begin{figure*}
    \centering
    \includegraphics[width=1\textwidth]{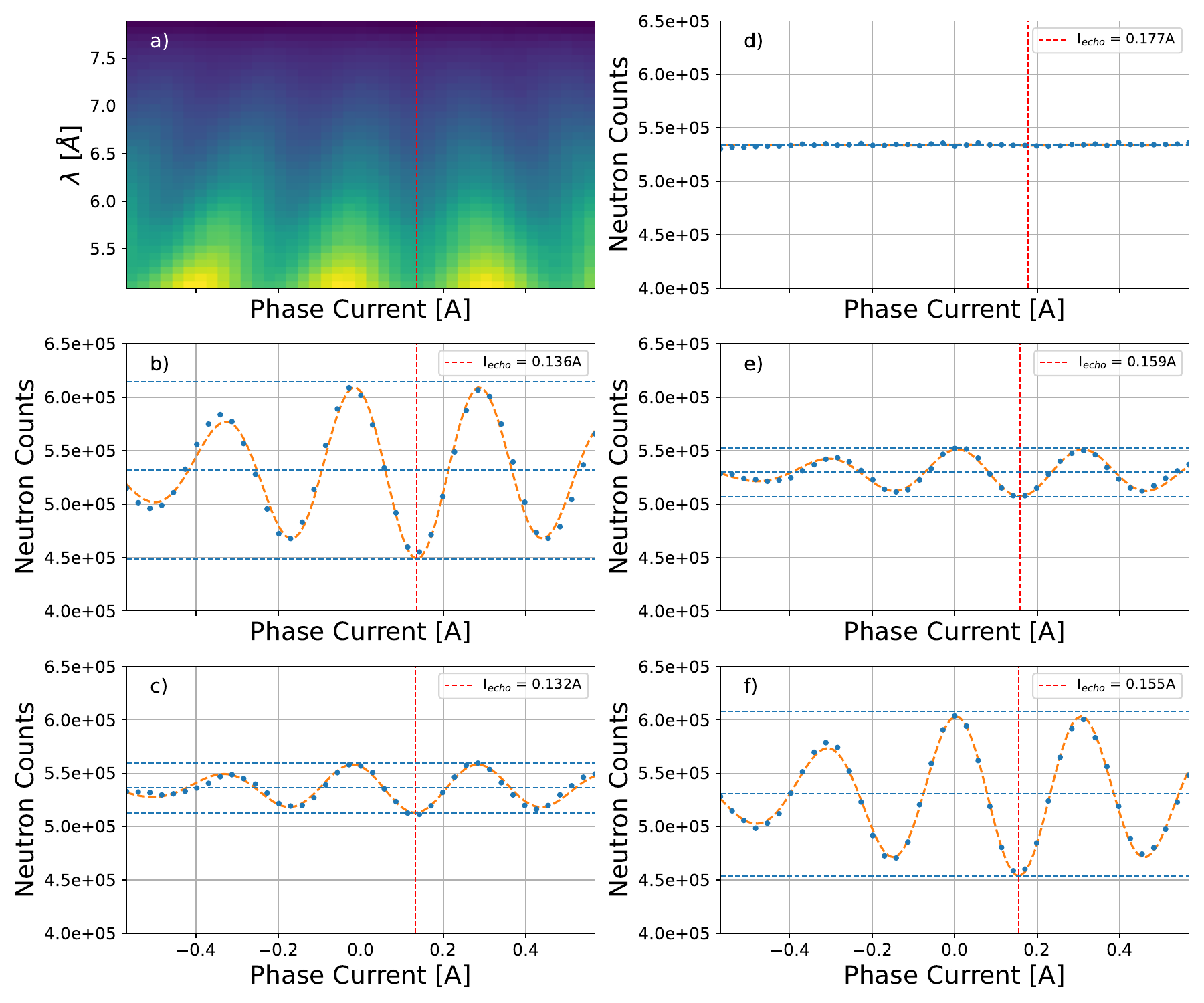}
    \caption{Plots of the neutron spin echo amplitude and phase in forward transmission as a function of the phase current in the spectrometer. Panel (a) is an example of a heat map of the intensity of neutron counts across the range of incident neutron wavelengths used as a function of the phase current for the sample at $T=230$ K. Panels (b) - (f) show a sum over the range of wavelengths used in the measurement at each value for the phase current at the following temperatures: 230 K (b), 240 K (c), 251 K (d), 260 K (e), and 270 K (f). The spin echo condition in paramagnetic mode occurs at the value of the phase current which minimizes this sum and for which the signal minimum is wavelength-independent, as indicated by the vertical dashed line in the plots.}
    \label{fig:paraflips_2}
\end{figure*}

Figure \ref{fig:paraflips_2} shows the amplitude and phase shift of the neutron spin echo envelope in forward transmission when operated in paramagnetic mode, which uses the sample itself as the $\pi$-flipper to form the spin-echo. In this case the central point of the echo occurs at an interference minimum. 

\begin{figure*}
    \centering
    \includegraphics[width=0.75\textwidth]{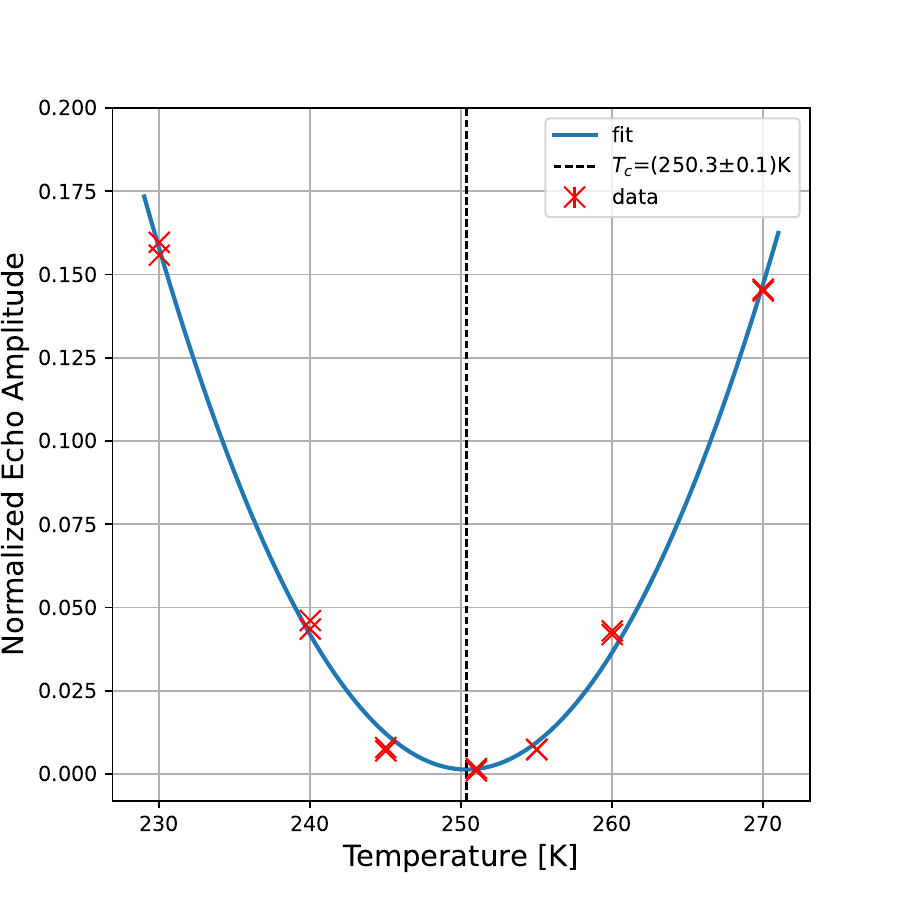}
    \caption{Plot of the centroid amplitude with a minimum around $T = 250.3$ K. The centroid amplitude is proportional to the magnetic moment of the sample with the neutron spin echo spectrometer operated in paramagnetic spin-flipping mode as shown in the data of Fig.~\ref{fig:paraflips_2}. The temperature where the spin echo amplitude is consistent with zero within statistical error is also consistent with the temperature $T_{comp}$ as determined by the SQUID external field measurements shown above and with compensation temperature measurements made in-situ.}
    \label{fig:NSE_Tc_2}
\end{figure*}

Figure \ref{fig:NSE_Tc_2} shows the centroid amplitude of the interference signal as a function of the temperature, with data points taken through both a heating and cooling cycle. This amplitude is that of the cosinusoidal signal shown in the plots displaying intensity as a function of phase current (Fig. \ref{fig:paraflips_2}). These measurements are consistent with the previous SQUID measurements and the in-situ measurements made with Hall probes and fluxgate magnetometers, showing an amplitude which vanishes within the statistical accuracy of the measurements around T = 250.3 K.

\begin{figure*}
    \centering
    \includegraphics[width=1\textwidth]{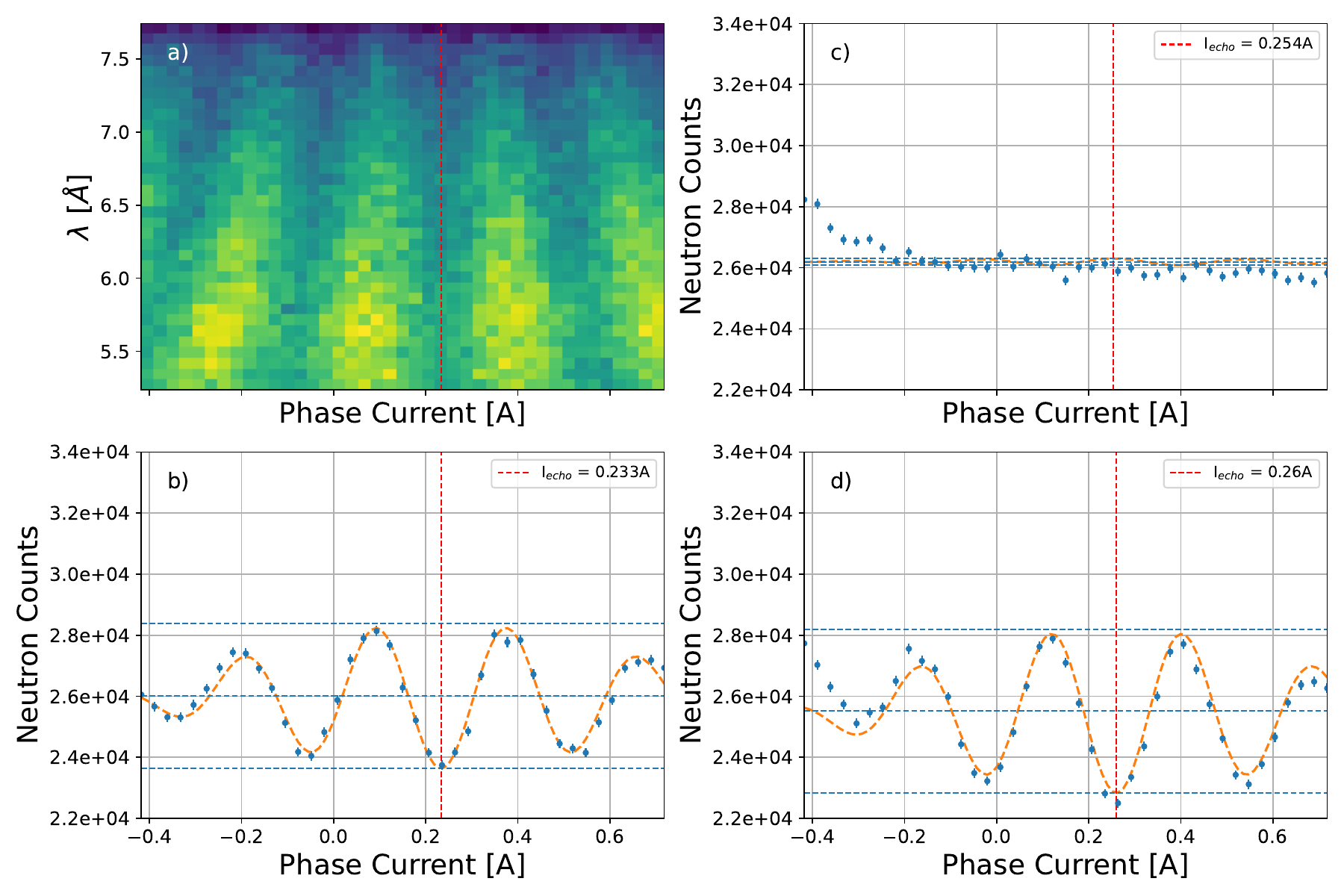}
    \caption{Plots of the neutron spin echo amplitude and phase as a function of the phase current in the spectrometer. In this case, the downstream arm of the spectrometer is set to an angle of 4 $\deg$ relative to the incident neutron momentum. Panel (a) is an example of a heat map of the intensity of neutron counts across the range of incident neutron wavelengths used as a function of the phase current for the sample at $T=240$ K. Panels (b) - (d) show a sum over the range of wavelengths used in the measurement at each value for the phase current at the following temperatures: 240 K (b), 251 K (c), and 260 K (d). The spin echo condition in paramagnetic mode occurs at the value of the phase current which minimizes this sum and for which the signal minimum is wavelength-independent, as indicted by the vertical dashed line in the plots.}
    \label{fig:paraflipsQne0}
\end{figure*}

Some data were also taken with the downstream arm of the spectrometer set to an angle of 4 $\deg$ relative to the incident neutron momentum. This latter setting corresponds to a range of momentum transfers $Q$ between 0.062 \AA$^{-1}$ and 0.088 \AA$^{-1}$ for the incident neutron energies of 1.3 meV and 3.2 meV selected by the upstream neutron choppers. The direction of the neutron momentum transfer for this spectrometer setting lies in the horizontal plane and points nearly normal to the incident neutron momentum direction. Figure~\ref{fig:paraflipsQne0} shows the neutron spin echo signal at temperatures of 230K, 251K, and 270K. 

These data were taken to check our understanding of the paramagnetic scattering from the sample. In paramagnetic neutron scattering, only the fluctuations in the internal magnetization normal to the momentum transfer direction $\vec{Q}$ cause neutron spin flip scattering and therefore contribute to the neutron spin echo signal. Since the sample was magnetized along the axis of the cylindrical sample which in turn is parallel to the incident neutron momentum and therefore normal to $\vec{Q}$, we expected to see a paramagnetic spin echo signal away from $T_{comp}$ but nothing at $T_{comp}$ as in the transmission data. The data indeed shows this behavior as expected, thereby confirming again that the internal magnetization in the sample away from  $T_{comp}$ points in the expected direction. Furthermore, the agreement between $T_{comp}$ as inferred from the $Q=0$ and $Q \neq 0$ data is also consistent within measurement precision with the absence of anisotropic components to the antiferromagnetic ordering from sample impurities. These data, however, possess lower statistical accuracy compared to our transmission data and therefore are not as sensitive to deviations away from $T_{comp}$.    

\section{Polarized Neutron Imaging Setup}
\label{sec:PolNeutronImagingSetup}

We used the MARS (Multimodal Advanced Radiography Station) neutron imaging instrument at HFIR, which we augmented to conduct polarized neutron spin rotation measurements.   A V-cavity polarizer in the slow neutron beam upstream of the target produced a high polarization transverse to the momentum. The neutron spin was adiabatically transported into a mu-metal magnetic shield and into the sample region. A sharp, nonadiabatic neutron spin transition after the sample region was established using a Forte-style V-coil followed by a longitudinal solenoidal magnetic field. This combination of magnetic fields adiabatically transported the component of the neutron spin which rotated about the neutron beam axis upon traversing the sample region and guided it into the neutron beam direction, where it was analyzed using a longitudinally-polarized $^{3}$He neutron spin filter based on spin-exchange optical pumping. By reversing the direction of the current in the Forte coil, this combination of fields reversed the direction of the transported rotated neutron spin component. The spin rotation angle could then be extracted from the asymmetry in the neutron intensity transmitted through the polarization analyzer, according to:

\begin{equation} 
PA\sin{\phi} = \frac{N_{+}-N_{-}}{N_{+}+N_{-}} ~.
\end{equation}

Here, $N_{+}$ and $N_{-}$ are the neutron transmission intensities through the polarization analyzer for the two directions in the Forte coil, $P$ is the neutron polarization ($\approx 0.95$), $A$ is the analyzing power of the neutron polarization analyzer ($\approx 0.86$), and $\phi$ is the neutron spin rotation angle. Figure \ref{fig:N_Spin_Components} shows a block diagram of the neutron spin transport components and Figure \ref{fig:N_Spin_Manipulations} illustrates the spin manipulations occurring at each stage.

\begin{figure}
    \centering
    \includegraphics[angle=-32,width=0.7\linewidth]{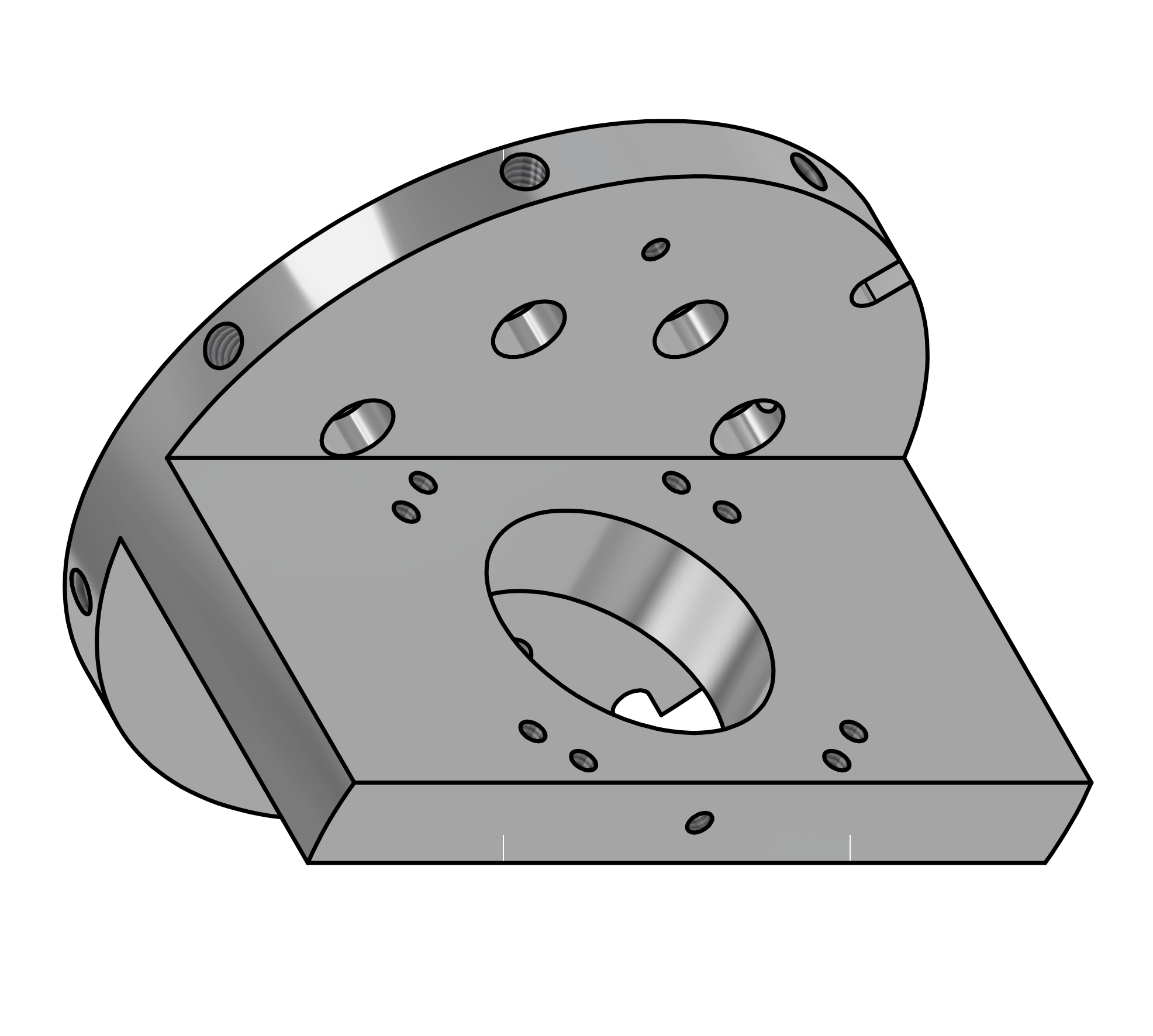}
    \caption{The single-cell sample holder used in the
2023 MARS experiment.}
    \label{fig:Single_Sample_Holder}
\end{figure}

\begin{figure*}
    \centering
    \subfloat[Block diagram of the neutron spin transport elements installed on the MARS beamline.]{
        \includegraphics[width=\linewidth]{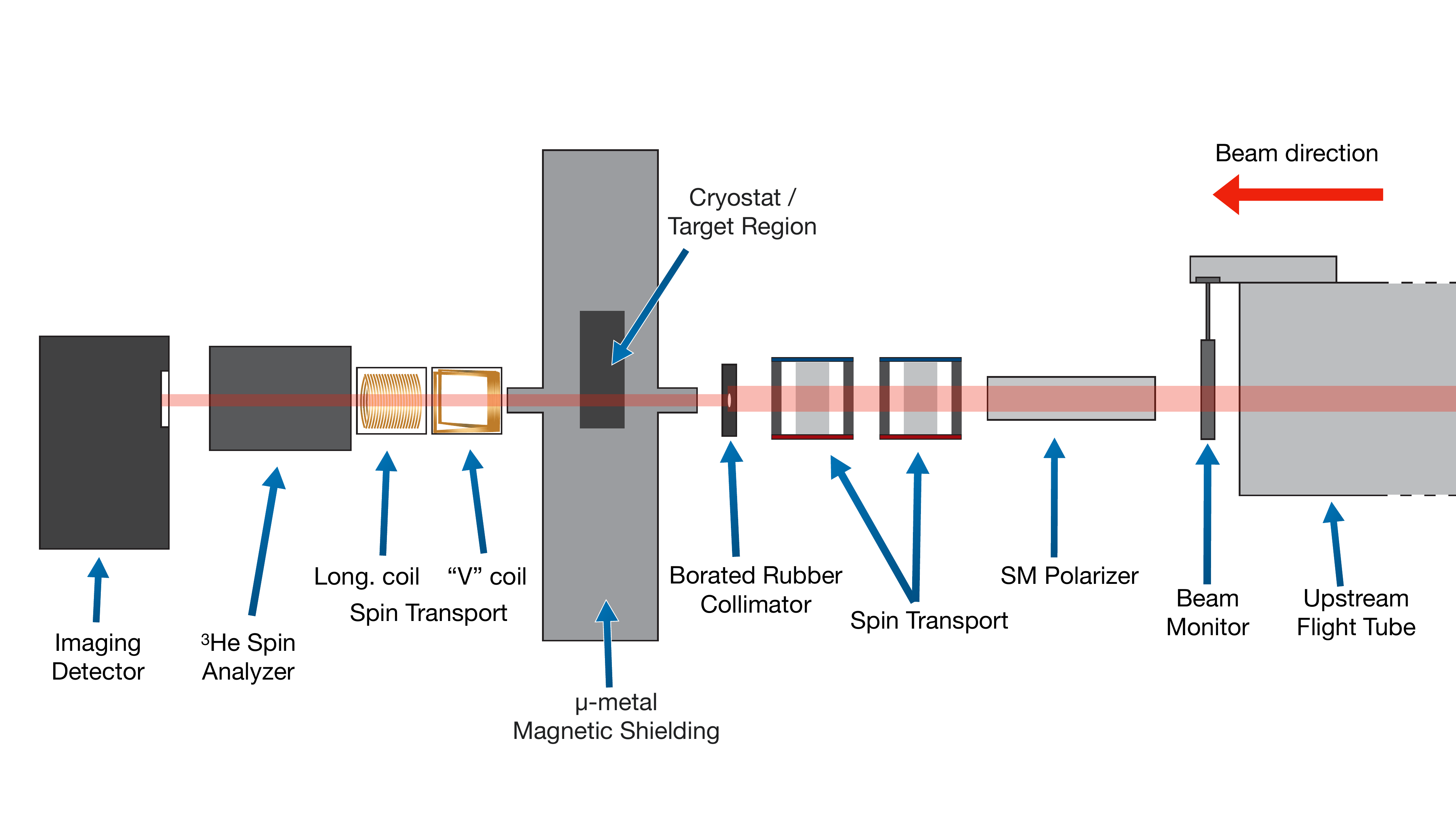}
        \label{fig:N_Spin_Components}
    }
    \hfill
    \subfloat[Block diagram of the neutron spin manipulations occurring at each stage along the beamline.]{
        \includegraphics[width=\linewidth]{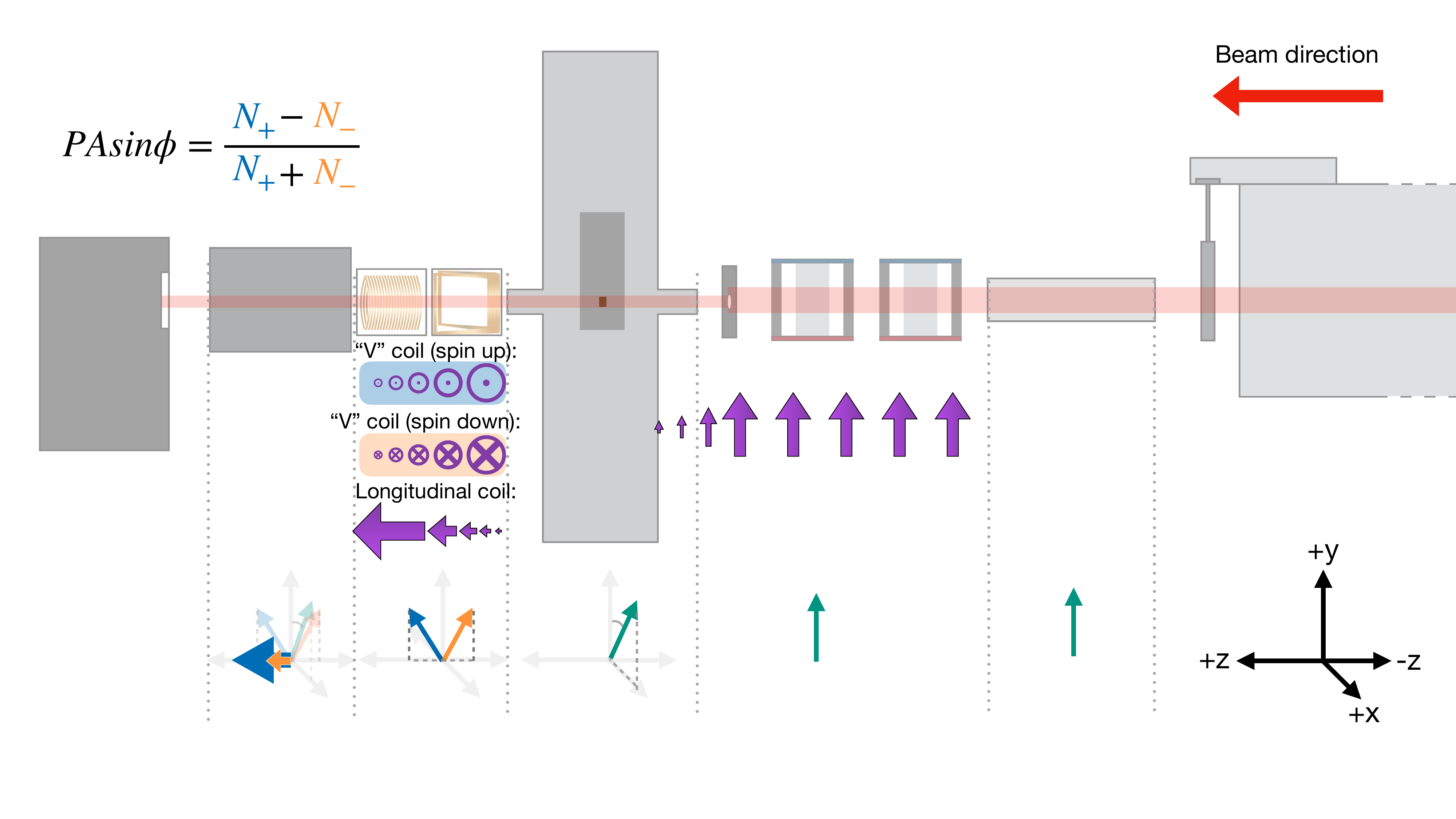}
        \label{fig:N_Spin_Manipulations}
    }
\caption{Block diagrams of the neutron spin transport elements installed in the MARS beamline. Each component is labeled in (\protect\subref{fig:N_Spin_Components}). The neutron spin manipulation is shown in (\protect\subref{fig:N_Spin_Manipulations}): purple arrows represent magnetic fields, green arrows represent the direction of neutron spin, and blue and orange arrows represent positive and negative rotations of the neutron spin, respectively.\vspace{5pt}}
\label{fig:Block_Diagram}
\end{figure*}

The TbIG sample was mounted with its axis of magnetization along the neutron beam direction (along the z axis in Figure \ref{fig:N_Spin_Manipulations}) so that any nonzero sample magnetization along the neutron beam axis would rotate the neutron spin in a direction that would be caught by the neutron polarimeter (in the xy plane as shown in Figure \ref{fig:N_Spin_Manipulations}). The neutron polarimeter was calibrated by applying a small longitudinal magnetic field just outside the entrance to the magnetic shield and confirming the expected neutron spin rotation angle from Larmor precession. 

Much of the same instrumentation from the SNS-NSE experiment was used to mount and cool the TbIG sample, with slight modifications. The sample holder design is shown in Figure \ref{fig:Single_Sample_Holder}. The TbIG sample was contained in a single-cell aluminum sample holder with sapphire windows placed in good thermal contact on both ends of the cylindrical sample and on the sample holder. Silicon diode thermometry determined the sample temperature to 10 mK precision. The cell was surrounded by an aluminum radiation shield and held in the vacuum chamber of a Cryo Industries RC102 cryostat with sapphire optical access windows. The cryostat was cooled using a slow stream of liquid nitrogen whose flow rate was set to bring the target temperature close enough to the temperatures of interest that the fine temperature control could be realized using a PID feedback system with a heater on the sample. Temperature stability of 50 mK was realized as data were collected in a 2.5 K band around $T_{comp}$ in 0.5 K steps. A fluxgate magnetometer just outside the cryostat vacuum near the cell was used to monitor the magnetic field inside the magnetic shield.   

The neutron detector was an imaging device with a maximum spatial resolution of $\approx$ 100 microns \cite{MARS23}. The spatial resolution of this imaging instrument was fine enough, in combination with the choices of neutron beam collimation made for this measurement, to identify macroscopic magnetic domains of size $\approx$ 1mm x 1mm from the neutron spin rotation angle measured at each temperature \cite{Dhiman17} if they exist. The results of a detailed analysis to search for any evidence of domain structure in our ferrimagnetic sample will be presented in a later paper. 

\begin{figure*}[hpbt!]
    \centering
    \includegraphics[clip, trim=0mm 15mm 0mm 8mm, width=2\columnwidth]{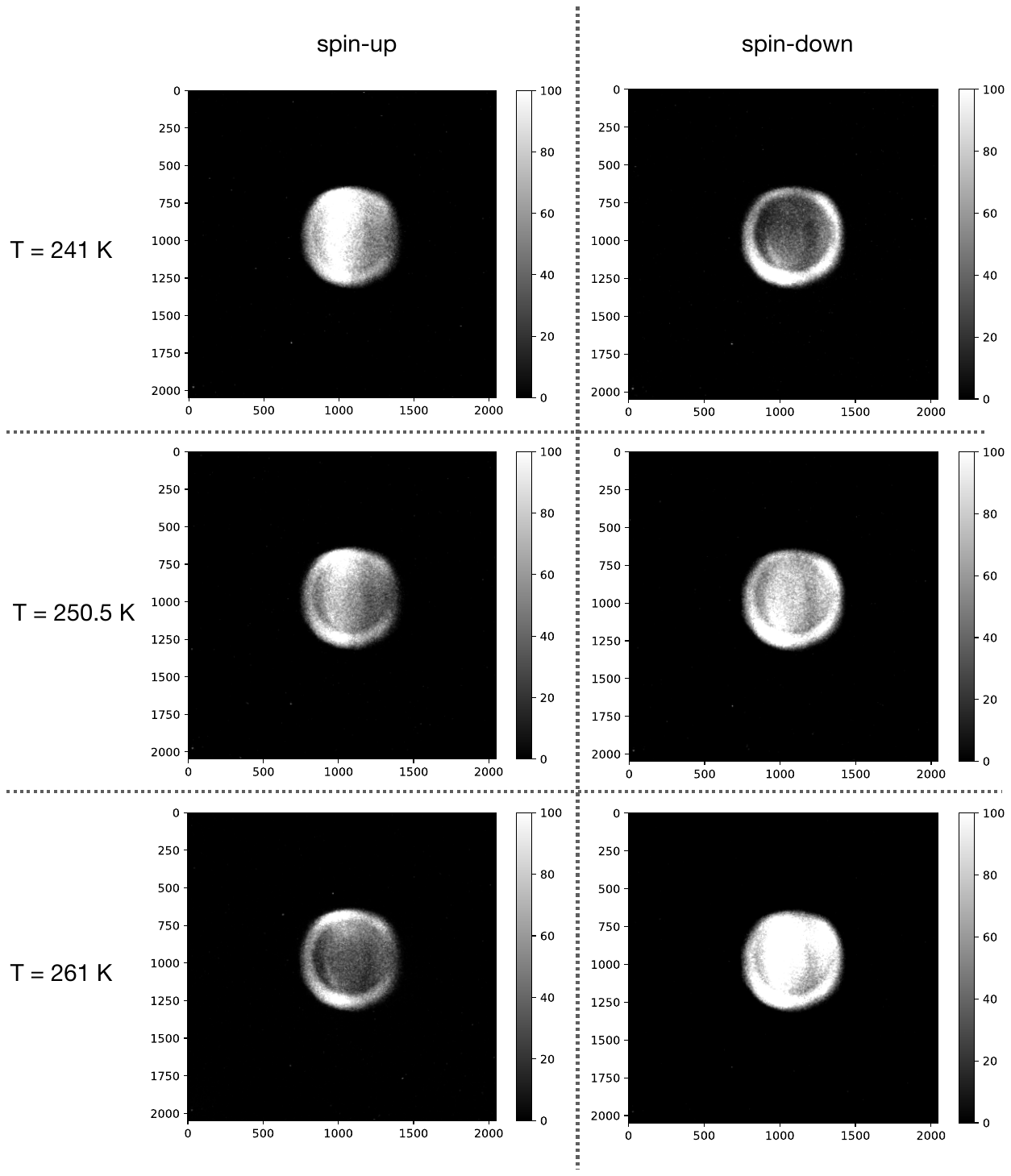}
    \caption{Radiographs of TbIG under different spin orientations and temperatures. The exposure time of each radiograph is 300 seconds with a 300 $\mu$m-thick RC Tritec $^{6}$LiF/ZNS:Cu (ratio 1/2) scintillator. The axes for each image refer to pixel number for a 2048x2048 pixel CCD camera detector. The same pixel intensity grayscale bar is used for all images. For a given spin state of neutrons in the beam (spin-up in the left column and spin-down in the right column of the plots), we see a transition from low to high (or high to low) intensity above and below the compensation temperature $T_{comp}$. Very close to $T_{comp}$, we see similar intensities implying roughly equal transmission for each spin state.}
    \label{fig:Radiographs}
\end{figure*}

\begin{figure*}
    \centering
    \includegraphics[width=1.7\columnwidth]{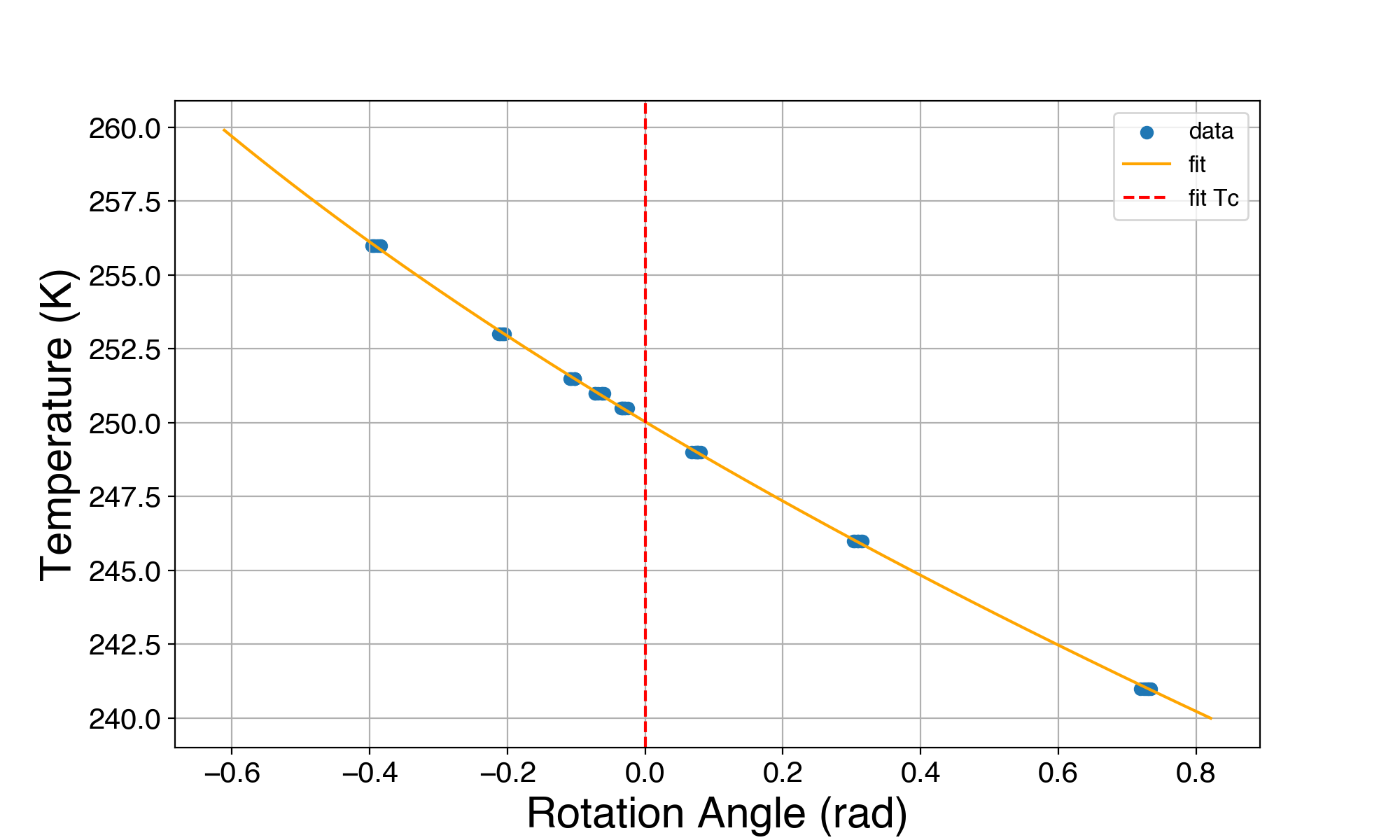}
    \caption{Plot of neutron spin rotation vs temperature and quadratic fit with x-intercept of $T_{comp}=250.03 \pm 0.03$ K.}
    \label{fig:NSR_Tc}
\end{figure*}

\begin{figure*}
    \centering
    \includegraphics[width=1.7\columnwidth]{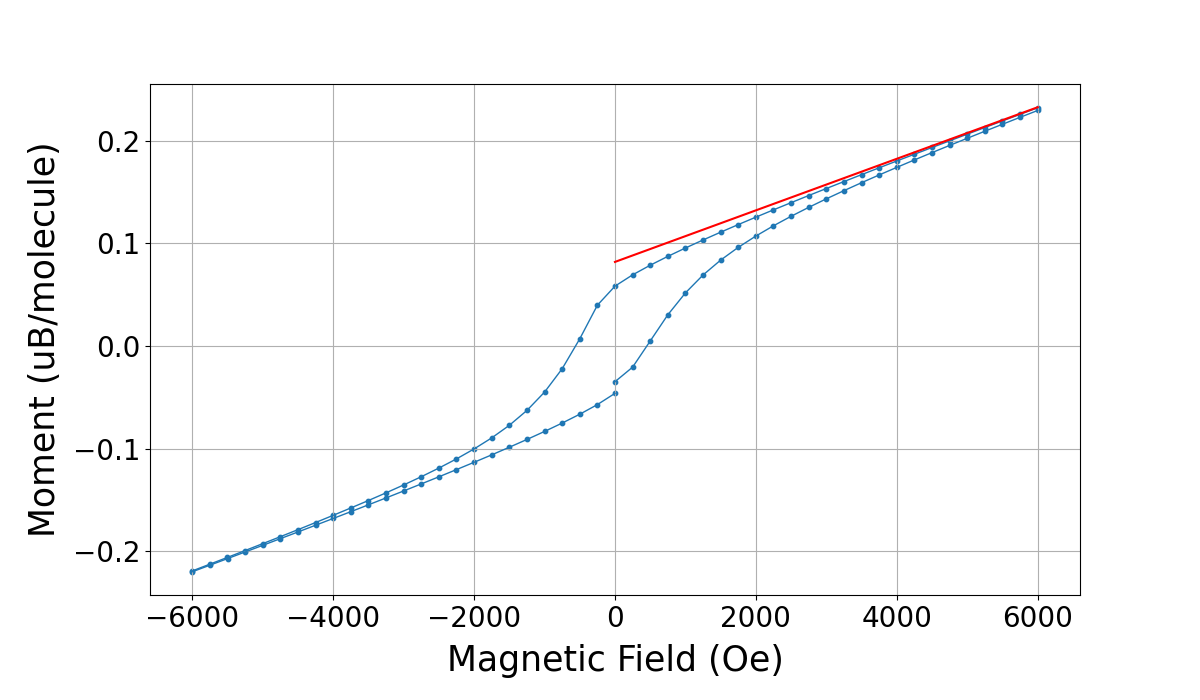}
    \caption{Moment vs. magnetic field hysteresis plot for a powder sample of TbIG at $260$~K. From this curve we see that at $T_{comp}+10$~K, the remnant magnetization is $0.058~\mu_B$/mol, and the value extrapolated back from the linear behavior at high fields is $0.08~\mu_B$/mol.}
    \label{fig:Tbig_Hysteresis_260}
\end{figure*}

\section{Polarized Neutron Spin Rotation Images}
\label{sec:PolNeutronSpinRotImages}

Figure \ref{fig:Radiographs} shows a sequence of neutron image pairs for the two different settings of the neutron polarimeter at three different temperatures: one pair above $T_{comp}$, one pair at $T_{comp}$ as inferred from the measurements of the external magnetic field and the neutron spin echo data, and one pair below $T_{comp}$. The neutron wavelength distribution used was that from the cold neutron source viewed by the MARS instrument as modified downstream by the upstream neutron guide and neutron supermirror polarizer; the mean neutron wavelength was $2.8$ \AA.  These sequences of images clearly show the internal magnetization inside the ferrimagnet vanishing at $T_{comp}$ and reversing sign on either side of $T_{comp}$. In addition, Figure \ref{fig:NSR_Tc} shows the measured neutron spin rotation angles at different temperatures. A quadratic fit to the data gives rotation angles of 0.82 and -0.62 radians at 10~K below and above $T_{comp}=250$~K, and 0 radians at $T=T_{comp}$, supporting the conclusion that the magnetization of the sample vanished at $T_{comp}$. 

The magnetization results can be compared to a molecular field model of ferrimagnetic garnets, which is described in detail in \cite{Leslie2014}  and is briefly summarized here. A three-sublattice model of TbIG is used with a Tb$^{3+}$ dodecahedral site denoted $c$, an Fe$^{3+}$ octahedral site denoted $a$, and an Fe$^{3+}$ tetrahedral site denoted $d$. As functions of temperature, the total magnetization per molecule of each sublattice is given by the starting magnetization multiplied by Brillouin functions $B_{J}$:

\begin{align}\label{eq:mag}
\begin{split}
    M_c(T)=M_c(0)B_{J_c}(x_c)\\
    M_a(T)=M_a(0)B_{J_a}(x_a)\\
    M_d(T)=M_d(0)B_{J_d}(x_d).
\end{split}
\end{align}

The Boltzmann energy ratios in eq.~\ref{eq:mag} are given by:

\begin{align}
\begin{split}
    x_c=\frac{g_cJ_c\mu_B}{k_BT} [N_{cc} M_c + N_{ac} M_a + N_{dc} M_d]\\
    x_a=\frac{g_aJ_a\mu_B}{k_BT} [N_{ac} M_c + N_{aa} M_a + N_{ad} M_d]\\
    x_d=\frac{g_dJ_d\mu_B}{k_BT} [N_{cd} M_c + N_{ad} M_a + N_{dd} M_d],
\end{split}
\end{align}
where the $N_{ij}$ are the molecular field coefficients. The coefficients (in mol/cm$^3$) are $N_{ac} = -4.2$, $N_{dc} = 6.5$, $N_{cc} = 0$, $N_{aa} = -65$, $N_{ad} = 97$, and $N_{dd} = -30.4$, taken from previous work \cite{Dionne09Book,LandoltBornstein1978:sm_lbs_978-3-540-37375-9_20}.

The molecular field model, corrected for small admixtures of orthoferrite and hematite (with moments described in \cite{Treves1965,Filho2014}), predicts magnetization values of 2589 A/m and -2328 A/m, at 10 K below and above $T_{comp}$, for the TbIG sample used in the NSE and imaging experiments. The predicted neutron spin rotation angles, based on the integration of the analytical expression of the longitudinal field on the axis of the sample with the above magnetizations, are 2.79 radians and -2.50 radians at these temperatures.

We note that the $N_{ij}$ in the model result from fits to the spontaneous moments obtained from high-field data extrapolated to zero applied field from the linear parts of garnet hysteresis curves, and represent the fully saturated moments.  The actual remanent magnetization we observe in our powder TbIG samples is smaller by a factor that varies somewhat with temperature.  Fig. 13 shows a hysteresis curve for a powder TbIG sample at 260 K (10 K above $T_{comp}$), measured in a SQUID magnetometer.  From the figure, the remanence is approximately 70\% of the extrapolated moment, which holds for 10 K below $T_{comp}$ as well. This 70\% correction appears to hold for our samples arbitrarily close to $T_{comp}$ (at room temperature the factor is closer to 50\%), and is sufficient to explain the discrepancy between the model and the SQUID remanence data.

In addition, the sample had not been polarized for at least 6 months prior to the imaging measurements at MARS (activation in the NSE beam precluded handling of the sample).  At polarization, just before the NSE measurements, the surface field was measured to be 16.3 Gauss.  Immediately prior to the MARS imaging run, we did manage a rudimentary measurement of the sample surface field with a hand-held Gaussmeter, which registered 5 Gauss approximately 2~mm above the surface. This suggests a demagnetization factor of about 1/3, which would affect all measurements correlated with the magnetization, including the spin rotation.

Applying the appropriate remanent scaling and demagnetization factors to the model predictions for the neutron imaging sample yields magnetizations of 617 A/m and -555 A/m, and associated neutron spin rotation angles of $0.67 \pm 0.12$ radians and $-0.60 \pm 0.11$ radians, at 10 K below and above $T_{comp}$.  Here, we have included error from the remanence correction (the extrapolation in Fig. 13), the surface field measurement (0.5 mm position error in the Gaussmeter probe), sample temperature (2 K), and we allow for 1\% variation in the $N_{ij}$.

An important prediction from the model is the (excess) electron spin density at $T_{comp}$.  Using the $N_{ij}$ above and applying the 70\% correction factor yields a value of $-0.37 \pm 0.02$ uB/molecule.  This prediction is also reduced by the sample purity, and depolarization factor affecting the spin rotation in the imaging measurements. The spin excess adjusted for these effects is $-0.11 \pm 0.02$ uB/molecule. 

\clearpage
\FloatBarrier

\section{Conclusions and Future Work}

We conclude from the evidence of all of the measurements described above that the internal magnetization of the ferrimagnet TbIG at and near its compensation temperature $T_{comp}$ over macroscopic internal distance scales larger than 1 mm behaves according to the usual macroscopic models of ferrimagnetism and possesses a vanishing magnetization at $T_{comp}$. 

We are conducting a more detailed analysis of the neutron spin rotation 2D image data to attempt to quantify the properties of any possible microscale ferrimagnetic domains that might be present in this material. This type of investigation is accessible through polarized neutron imaging but not to any external probe of the magnetization. 

With the same sample used in this work, we also conducted measurements at $T_{comp}$ in which we repeatedly rotated the sample 180 degrees about a vertical axis, thereby reversing the direction of the nonzero net electron spin polarization in the material, which was set by the prior sample magnetization to point along the cylinder axis and therefore along the neutron beam momentum. If a spin-dependent exotic interaction exists proportional to $\vec{s_{n}} \cdot \vec{s_{e}}$ where $\vec{s_{n}}$ is the neutron spin and $\vec{s_{e}}$ is the electron spin, then the transversely-polarized neutron spin would rotate about the neutron beam axis with an angle that reverses in sign upon reversal of $\vec{s_{e}}$. We will report the results of this analysis in a later work.

{\bf Acknowledgements}

The authors would like to acknowledge the contributions of Maren Pink at IU's Molecular Structure Center for her help with the x-ray diffractometry. The diffractometry was primarily conducted on a Bruker Venture D8 diffractometer. Support for the acquisition of the Bruker Venture D8 diffractometer through the Major Scientific Research Equipment Fund from the President of Indiana University and the Office of the Vice President for Research is gratefully acknowledged. Additional diffractometry was conducted on a PANalytical Empyrean multipurpose diffractometer, which is supported by National Science Foundation grant number NSF/CHE-1048613.  K. N. Lopez, M. Luxnat, M. Van Meter, S. Samiei, and W. M. Snow acknowledge support from US National Science Foundation (NSF) grants PHY-1913789 and PHY-2209481 and the Indiana University Center for Spacetime Symmetries. K. N. Lopez also acknowledges the support of the National GEM Consortium and the Indiana Space Grant Consortium. M. Sarsour and T. Mulkey acknowledge support from US Department of Energy grant DE-SC0010443. J. C. Long and B. Hill acknowledge support from US National Science Foundation (NSF) grant PHY-1707986. The authors acknowledge the use of facilities and instrumentation supported by NSF through the University of Illinois Materials Research Science and Engineering Center DMR-2309037. This research used resources at the High Flux Isotope Reactor and the Spallation Neutron Source, both DOE Office of Science User Facilities operated by the Oak Ridge National Laboratory. 

\bibliography{references} 

\end{document}